\documentclass[aps,reprint,pre,twocolumn,floatfix,amsmath,amssymb]{revtex4}
\usepackage[dvipdfm]{graphicx}
\usepackage{amsmath}
\usepackage{amssymb}
\usepackage{color}
\usepackage{dcolumn}
\begin{document}
\preprint{}
\title{
Microrheology of active suspensions
}
\author{Takahiro Kanazawa$^1$ and Akira Furukawa$^2$}
\thanks{furu@iis.u-tokyo.ac.jp}
\affiliation{$^1$Department of Physics, University of Tokyo, Bunkyo-ku, Tokyo 113-0033, Japan\\ $^2$Institute of Industrial Science, 
University of Tokyo, Meguro-ku, Tokyo 153-8505, Japan}
\date{\today}
\begin{abstract}
We study the microrheology of active suspensions through direct hydrodynamic simulations using model pusher-like microswimmers. We demonstrate that the friction coefficient of a probe particle is notably reduced by hydrodynamic interactions (HIs) among a moving probe and the swimmers.  When a swimmer approaches a probe from the rear (front) side, the repulsive HIs between them are weakened (intensified), which results in a slight front rear asymmetry in swimmer orientation distribution around the probe, creating a significant additional net driving force acting on the probe from the rear side. 
The present drag-reduction mechanism qualitatively differs from that of the viscosity-reduction observed in sheared bulk systems and depends on probing details. 
This study provides insights into our fundamental knowledge of hydrodynamic effects in active suspensions and serves as a practical example illuminating distinctions between micro- and macrorheology measurements. 
\end{abstract}

\maketitle

\section{Introduction}
In active suspensions, the intrinsic activity of swimming particles leads to distinctive collective behaviors and transport/rheological properties deviating from those observed in passive particle suspensions \cite{Review_active1,Review_active2,Review_active3}. 
A striking example is anomalous rheology \cite{Hatwalne,Sokolov,Gachelin,Lopez,Liu,PNAS2020,Rafai,Marchetti,Review3,Saintillan1,Ishikawa_Pedray,
Cates,Haines,Giomi,Ryan,Moradi_Najafi,Nechtel_Khair,Takatori_Brady,Hayano_Furukawa,Bayram}. In particular, in suspensions of pusher-like microswimmers, such as {\it E. coli}, the viscosity significantly decreases \cite{Sokolov,Gachelin,Lopez,Liu,PNAS2020}, frequently establishing zero or negative viscosity states \cite{Lopez,PNAS2020,Marchetti}. The underlying mechanism behind such anomalous rheology involves weak orientational order along the extension axis of the externally applied flow field, which could be attributed to rotational diffusivities and/or hydrodynamic interactions (HIs) \cite{Saintillan1,Review3,Ryan,Haines,Hayano_Furukawa}.  
The active dipolar forces with the orientational order intensify the mean flow, reducing the resistive stress required to drive the external flow and consequently diminishing the viscosity \cite{Hatwalne}. The anomalous viscosity reduction in active suspensions contrasts with the viscosity behavior of dilute suspensions of passive particles, which is well described by the Einstein viscosity formula \cite{Landau_LifshitzB}. 

The local viscosity or viscoelastic properties of active suspensions are of great interest; probing rheological properties at the $\mu$m level can provide further insights into the underlying mechanisms and enable a more detailed characterization of the anomalous rheology. 
When probing smaller-scale rheological properties of complex fluids or soft materials, microrheology measurements are considered powerful tools (see recent reviews \cite{Squires_Mason_review,MicrorheologyB,Zia_review} and the references therein). 
By tracking the motions of small probes suspended in fluids, typically at a $\mu$m scale, microrheology allows for measuring viscoelastic properties across a wider range of temporal scales than conventional macrorheology techniques. 
This finer resolution in both time and length scales may offer a more comprehensive understanding of the material's rheological properties. 
However, microrheology still faces several unresolved issues. One such issue is whether the observations made using microrheology accurately reflect the actual local rheological properties of the material in the absence of probes.
The interactions among probes and suspended constituents or inner structures can significantly influence the motion of the probes and potentially alter the local  rheological properties around the probe. Understanding and accounting for these effects are crucial for accurately interpreting microrheology measurements and relating them to a material's intrinsic viscoelastic properties.

Microrheological investigations on active suspensions have provided various intriguing results \cite{Wu,Chen,Leptos,Molina-Nakayama-Yamamoto,Foffano,Foffano2,Burkholder1,Burkholder2,Peng,Knezevic1,Knezevic2,Loisy,Reichhardt}. 
A simulation study by Foffano {\it et al}. \cite{Foffano,Foffano2} demonstrated that a probe particle can experience a negative viscosity in active nematics composing contractile-puller-type swimmers. More recent numerical studies \cite{Burkholder1,Burkholder2,Peng,Knezevic2} have predicted that the friction coefficient of a probe can be reduced in active Brownian particle baths without HIs, but not to a level smaller than that defined in the absence of active/passive Brownian particles, indicating that the measured viscosity is larger than the solvent viscosity.  However, as described above, in bulk rheological experiments of extensile-pusher-type swimmers, it has been observed that the measured viscosity can be lower than the solvent viscosity and even approach zero. 
This raises a question of whether such a phenomenon can be replicated in microrheology measurements when HIs are taken into account. 
If so, this would prompt further investigation into whether the reduction of the friction coefficient of a probe shares a similar mechanism to the viscosity reduction observed in macrorheology.

\section{Simulation methods}
\begin{figure*}[bt] 
\centering
\includegraphics[width=16cm]{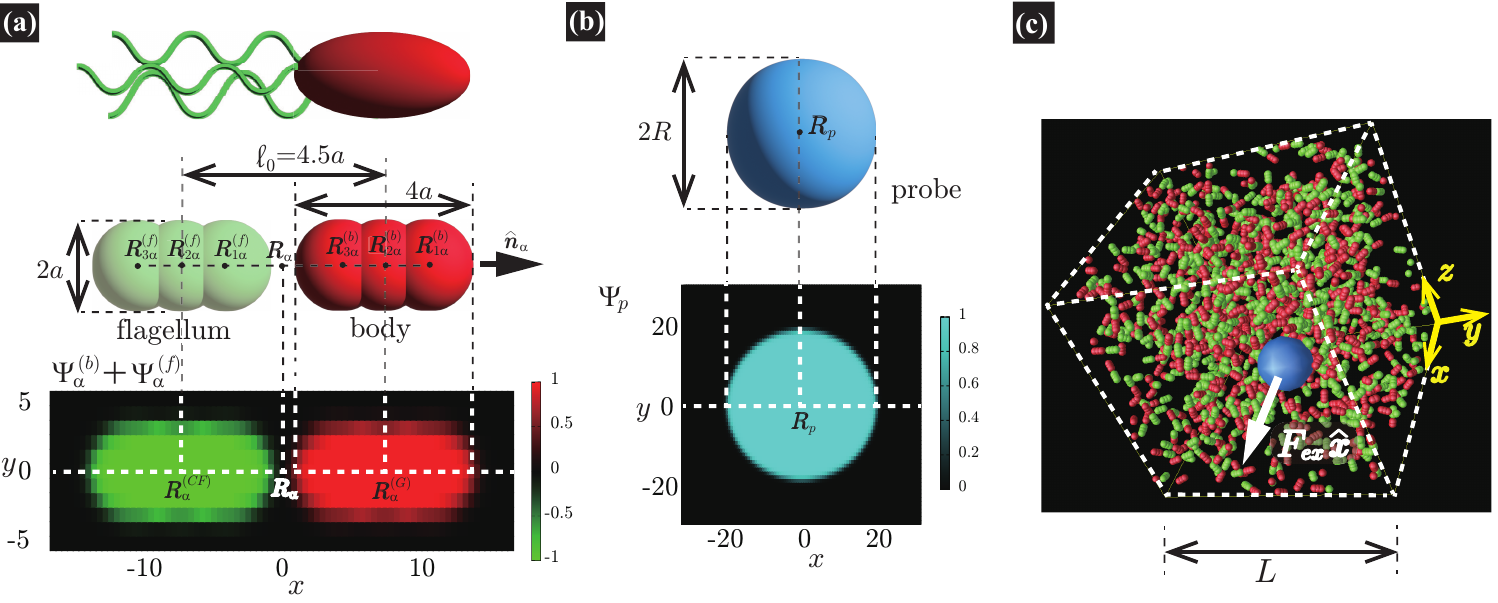}
\caption{(a) Our model swimmer comprises body and flagellum parts with symmetric shapes. Each part is constituted by a superposition of three spheres with radius $a$. We assume that a force $F_A{\hat {\mbox{\boldmath$n$}}}_\alpha$ is exerted on the body, while $-F_A{\hat {\mbox{\boldmath$n$}}}_\alpha$ is directly exerted on the solvent through the flagellum part, with ${\hat {\mbox{\boldmath$n$}}}_\alpha$ being the orientation of the $\alpha$-th swimmer. These forces constitute a force dipole of magnitude $F_A \ell_0$, with $\ell_0$ being the characteristic swimmer's length, which for the present model is given as the separation distance between the body and flagellum centers. In this study, to incorporate the present model swimmer into the SPM, the body and flagellum parts are represented through field variables, $\Psi_\alpha^{(b)}({{\mbox{\boldmath$r$}}})$ and $\Psi_\alpha^{(f)}({{\mbox{\boldmath$r$}}})$, respectively. In the bottom panel, we plot $\Psi_\alpha^{(b)}({{\mbox{\boldmath$r$}}})+\Psi_\alpha^{(f)}({{\mbox{\boldmath$r$}}})$ in the $xy$-plane, where both ${{\mbox{\boldmath$R$}}}_{\alpha}^{(G)}=(2.25a,0,0)$ and  ${{\mbox{\boldmath$R$}}}_{\alpha}^{(CF)}=(-2.25a,0,0)$ are included.   
(b)  In our microrheology simulation, a single probe particle with radius $R$ is immersed in a fluid. 
The probe particle is also described by the field variable $\Psi_{p}({{\mbox{\boldmath$r$}}})$.  
In the bottom panel, we plot $\Psi_{p}({{\mbox{\boldmath$r$}}})$ representing the probe particle in the $xy$-plane, where the probe center is set to ${{\mbox{\boldmath$R$}}}_{p}=(0,0,0)$.  
In (a) and (b), the discretized mesh size $h$ is the same as that used in practical simulations ($h=0.3125a$ and $\xi=0.5h$). Here, $\xi$ is the interface thickness controlling the degree of smoothness of $\Psi_\alpha^{(b)}({{\mbox{\boldmath$r$}}})$, $\Psi_\alpha^{(f)}({{\mbox{\boldmath$r$}}})$, 
and $\Psi_{p}({{\mbox{\boldmath$r$}}})$. 
(c) A single probe particle immersed in a fluid is dragged by a constant force $F_{ex}$ along the $x$-direction. The periodic boundary conditions are imposed in all directions with the linear dimension $L$.  }
\label{Fig1}
\end{figure*}

For the present purpose, we conduct direct hydrodynamic simulations using a model of active suspensions comprised of $N$ rod-like dumbbell swimmers. 
Our model swimmer, shown in Fig. 1(a), is essentially the same as the one employed in our previous investigation of macrorheology simulations \cite{Hayano_Furukawa}. 
In this model, each swimmer consists of a body and a flagellum: the body is treated as a rigid-body particle, while the flagellum is considered a massless ``phantom'' particle that simply follows the body's motions. This treatment maintains the relative position of the body and flagellum parts. 
For the $\alpha$-th swimmer ($\alpha=1,\cdots, N$), we assume that a force $F_A{\hat {\mbox{\boldmath$n$}}}_{\alpha}$ acting on the (front) body is exerted by the (rear) flagellum and that the flagellum also exerts the force $-F_A{\hat {\mbox{\boldmath$n$}}}_{\alpha}$ directly on the solvent fluid. 
Here, ${\hat {\mbox{\boldmath$n$}}}_{\alpha}$ is the direction of the $\alpha$-th swimmer, and these forces compose a dipolar force of magnitude $F_A\ell_0$, with $\ell_0$ being the characteristic swimmer's length [see Fig. 1(a)]. 
Such a force-prescribed particle model emulates rod-like pusher-type microorganisms such as {\it E. coli}, as initially proposed in Refs. \cite{Graham1,Graham2} 
and employed in subsequent studies \cite{Saintillan-ShelleyP,Haines,Ryan,Gyrya,Decoene,Furukawa_Marenduzzo_Cates,Hayano_Furukawa}. 

As illustrated in Fig. \ref{Fig1}(a), the body and flagellum parts are assumed to have the same shape and are each described by a superposition of three spheres with a common radius $a$. 
The spheres composing the body are located at the positions  
${{\mbox{\boldmath$R$}}}_{i,\alpha}^{(b)}={{\mbox{\boldmath$R$}}}_{\alpha}^{(G)}+(2-i)a{\hat {\mbox{\boldmath$n$}}}_{\alpha}$ ($i=1,2,3$), where ${{\mbox{\boldmath$R$}}}_{\alpha}^{(G)}$ is the $\alpha$-th swimmer's center-of-mass position. 
Similarly, the spheres composing the flagellum part are located at ${{\mbox{\boldmath$R$}}}_{i,\alpha}^{(f)}={{\mbox{\boldmath$R$}}}_{\alpha}^{(CF)}+(2-i)a{\hat {\mbox{\boldmath$n$}}}_{\alpha}$ ($i=1,2,3$), where ${{\mbox{\boldmath$R$}}}_{\alpha}^{(CF)}={{\mbox{\boldmath$R$}}}_{\alpha}^{(G)}-4.5a{\hat {\mbox{\boldmath$n$}}}_{\alpha}={{\mbox{\boldmath$R$}}}_{\alpha}^{(G)}-\ell_0{\hat {\mbox{\boldmath$n$}}}_{\alpha}$ is the position of the center of the flagellum. 
The shape of the present model swimmer shows the head-tail symmetry, and the mid-point is thus given by ${{\mbox{\boldmath$R$}}}_{\alpha}=({{\mbox{\boldmath$R$}}}_{\alpha}^{(G)}+{{\mbox{\boldmath$R$}}}_{\alpha}^{(CF)})/2$. 
Although arbitrary shapes of swimmers with an imposed head tail asymmetry can be composed, we can obtain qualitatively the same results as long as these swimmers have rod-like forms with the prescribed force dipoles.

In our simulations, we use the smoothed profile method (SPM) \cite{SPM,SPM2,SPM3} to accommodate many-body hydrodynamic interactions (HIs) among the constituent swimmers. 
In Ref. \cite{SPM3}, 
it was found that the SPM can quantitatively reproduce far-field and intermediate-field aspects of HIs, whereas near-field HIs are slightly underestimated at closer distances. 
Furthermore, like many other methods \cite{LB1,LB2,FPD,FPD2,DPD,MPC}, the SPM cannot also resolve the singular lubrication forces. For more details of the qualitative evaluations on the SPM, please refer to Refs. \cite{SPM2,SPM3}.  

Here, we describe a detailed scheme to simulate the present model swimmer system, essentially the same as that used in a previous study \cite{Hayano_Furukawa}. 
The body and flagellum parts of a swimmer are represented through the field variables $\Psi_\alpha^{(b)}({{\mbox{\boldmath$r$}}})$ and $\Psi_\alpha^{(f)}({{\mbox{\boldmath$r$}}})$, respectively:  
\begin{eqnarray}
\Psi_\alpha^{(b)}({{\mbox{\boldmath$r$}}}) = {\rm min}\biggl\{ \sum_{i=1}^{3}\psi [{{\mbox{\boldmath$r$}}}, {{\mbox{\boldmath$R$}}}_{i,\alpha}^{(b)};a],1\biggr\} \label{profileB}
\end{eqnarray}
and 
\begin{eqnarray}
\Psi_\alpha^{(f)}({{\mbox{\boldmath$r$}}}) = {\rm max}\biggl\{-\sum_{i=1}^{3}\psi [{{\mbox{\boldmath$r$}}}, {{\mbox{\boldmath$R$}}}_{i,\alpha}^{(f)};a],-1 \biggr\}.  \label{profileF}
\end{eqnarray} 
In this study, we adopt the following function to $\psi$ as 
\begin{eqnarray}
\psi [{{\mbox{\boldmath$r$}}}, {{\mbox{\boldmath$R$}}}_{i,\alpha}^{(\mu)};a]=\dfrac{1}{2}\biggl\{\tanh\biggl[\dfrac{1}{\xi}(a-|{{\mbox{\boldmath$r$}}}-{{\mbox{\boldmath$R$}}}_{i,\alpha}^{(\mu)}|)\biggr] +1 \biggr\},  
\end{eqnarray} 
where $\mu=b,f$ and $\xi$ is the interface thickness controlling the degree of smoothness. 
In Fig. \ref{Fig1}(a), we show the cross section of the model swimmer described by $\Psi_\alpha^{(b)}({{\mbox{\boldmath$r$}}})$ and $\Psi_\alpha^{(f)}({{\mbox{\boldmath$r$}}})$, including both ${{\mbox{\boldmath$R$}}}_{\alpha}^{(G)}$ and  ${{\mbox{\boldmath$R$}}}_{\alpha}^{(CF)}$ in the same plane.

In our microrheology simulation, a single probe particle (radius $R$) is immersed in a fluid, which is dragged by a constant force $F_{ex}$ along the $x$-direction. 
Similarly to the swimmers, as shown in Fig. \ref{Fig1}(b), the probe particle is also described by the field variable $\Psi_{p}({{\mbox{\boldmath$r$}}})$ as  
\begin{eqnarray}
\Psi_{p}({{\mbox{\boldmath$r$}}}) &=& \psi [{{\mbox{\boldmath$r$}}}, {{\mbox{\boldmath$R$}}}_{p};R], \label{profileP}
\end{eqnarray}
where ${{\mbox{\boldmath$R$}}}_{p}$ is the position of the probe sphere's center and $R$ is the probe radius.

The working equations for the velocity field  ${{\mbox{\boldmath$v$}}}({{\mbox{\boldmath$r$}}},t)$ are given as 
\begin{eqnarray}
\rho\biggl(\dfrac{\partial}{\partial t}+ {{\mbox{\boldmath$v$}}}\cdot {{\mbox{\boldmath$\nabla$}}}\biggr){{\mbox{\boldmath$v$}}} &=&  {{\mbox{\boldmath$\nabla$}}}\cdot {\stackrel{\leftrightarrow}{\mbox{\boldmath$\Sigma$}}_{vis}}-{{\mbox{\boldmath$\nabla$}}}p+ {{\mbox{\boldmath$f$}}}_{H}+ {{\mbox{\boldmath$f$}}}_{A}^{(f)},   
\label{Navier_Stokes} \\ 
{\stackrel{\leftrightarrow}{\mbox{\boldmath$\Sigma$}}_{vis}} &=&  \eta_s \bigl[{{\mbox{\boldmath$\nabla$}}} {{\mbox{\boldmath$v$}}}+ ({{\mbox{\boldmath$\nabla$}}} {{\mbox{\boldmath$v$}}})^\dagger\bigr], \label{stress_tensor} \\
 {{\mbox{\boldmath$\nabla$}}}\cdot {{\mbox{\boldmath$v$}}}&=&0. \label{incompressibility}
\end{eqnarray}
Equation (\ref{Navier_Stokes}) is the usual Navier-Stokes equation \cite{Landau_LifshitzB}. 
Here, $\rho$ is the solvent mass density, ${\stackrel{\leftrightarrow}{\mbox{\boldmath$\Sigma$}}_{vis}}$, given as Eq. (\ref{stress_tensor}), is the viscous stress tensor with $\eta_s$ being the solvent viscosity, and the hydrostatic pressure $p$ is determined by the incompressibility condition, Eq.  (\ref{incompressibility}). 
In addition, ${{\mbox{\boldmath$f$}}}_{H}$ is the body force required to satisfy the rigid-body condition for the swimmer's body and probe particle regions, and 
${{\mbox{\boldmath$f$}}}_{A}^{(f)}$ is the active force directly exerted by the flagellum part to the fluid:  
\begin{eqnarray}
{{\mbox{\boldmath$f$}}}_{A}^{(f)}({{\mbox{\boldmath$r$}}})= \dfrac{1}{{\mathcal V}_{\alpha}^{(f)}}\sum_{\alpha=1}^{N}\Psi_\alpha^{(f)}({{\mbox{\boldmath$r$}}}){\hat{\mbox{\boldmath$n$}}}_{\alpha}F_A,  \label{active_forceF}
\end{eqnarray}
where ${\mathcal V}_{\alpha}^{(f)}=-\int {\rm d}{{\mbox{\boldmath$r$}}} \Psi_\alpha^{(f)}({{\mbox{\boldmath$r$}}})$ is the volume of the flagellum part. 
In addition, the volume of the body part is given as ${\mathcal V}_{\alpha}^{(b)}=\int {\rm d}{{\mbox{\boldmath$r$}}} \Psi_\alpha^{(b)}({{\mbox{\boldmath$r$}}})$. 
In this study, because the shapes of the body and flagellum parts are assumed to be the same, ${\mathcal V}_{\alpha}^{(b)}={\mathcal V}_{\alpha}^{(f)}$.

For the model swimmers, the equations of motion for the center-of-mass velocity ${{\mbox{\boldmath$V$}}}_\alpha^{(G)}$ and the angular velocity with respect to the center-of-mass ${{\mbox{\boldmath$\Omega$}}}_\alpha^{(G)}$ are 
\begin{eqnarray}
M_\alpha \dfrac{d{{\mbox{\boldmath$V$}}}_\alpha^{(G)}}{dt} &=& {{\mbox{\boldmath$F$}}}_{\alpha, {H}} + {{\mbox{\boldmath$F$}}}_{\alpha, {int}}+ {{\mbox{\boldmath$F$}}}_{\alpha, {A}}^{(b)},  \label{VG} 
\\ 
{\stackrel{\leftrightarrow}{\mbox{\boldmath$I$}}}_{\alpha}\cdot 
\dfrac{d{{\mbox{\boldmath$\Omega$}}}_\alpha^{(G)}}{dt} &=&  {{\mbox{\boldmath$N$}}}_{\alpha, {H}} + {{\mbox{\boldmath$N$}}}_{\alpha, {int}},   \label{OG}
\end{eqnarray}
where
\begin{eqnarray}
M_{\alpha} = \rho{\mathcal V}_\alpha^{(b)} \label{mass} 
\end{eqnarray} 
and 
\begin{eqnarray}
{\stackrel{\leftrightarrow}{\mbox{\boldmath$I$}}}_{\alpha} &=&   \int {\rm d}{{\mbox{\boldmath$r$}}} \rho\Psi_\alpha^{(b)}({{\mbox{\boldmath$r$}}})\biggl[|\Delta{{\mbox{\boldmath$r$}}}_\alpha |^2 {\stackrel{\leftrightarrow}{\mbox{\boldmath$\delta$}}} - \Delta{{\mbox{\boldmath$r$}}}_\alpha \Delta{{\mbox{\boldmath$r$}}}_\alpha \biggr]
\end{eqnarray}
are the mass and the moment of inertia of the $\alpha$-th swimmer's body, respectively. 
Here, ${\stackrel{\leftrightarrow}{\mbox{\boldmath$\delta$}}}$ is the unit tensor and $\Delta{{\mbox{\boldmath$r$}}}_\alpha= {{\mbox{\boldmath$r$}}}-{{\mbox{\boldmath$R$}}}_\alpha^{(G)}$. In this study, the swimmer's density is assumed to be the same as the solvent density. 
In Eqs. (\ref{VG}) and (\ref{OG}), ${{\mbox{\boldmath$F$}}}_{\alpha, {H}}$ and ${{\mbox{\boldmath$N$}}}_{\alpha, {H}}$ are the force and torque exerted on the $\alpha$-th swimmer due to HIs. 
The explicit forms of ${{\mbox{\boldmath$F$}}}_{\alpha, {H}}$, ${{\mbox{\boldmath$N$}}}_{\alpha, {H}}$, and the body force  ${{\mbox{\boldmath$f$}}}_{H}$ are given in the Appendix A. 
${{\mbox{\boldmath$F$}}}_{\alpha, {int}}$ and ${{\mbox{\boldmath$N$}}}_{\alpha, {int}}$ are the force and torque acting on the $\alpha$-th swimmer's body, respectively, due to the particle-particle and particle probe potential interactions: 
\begin{eqnarray}
{{\mbox{\boldmath$F$}}}_{\alpha, {int}} &=& -\sum_{\beta\ne \alpha}\sum_{i,\mu\in \alpha}\sum_{j,\nu\in \beta} \dfrac{\partial}{\partial {{\mbox{\boldmath$R$}}}_{i,\alpha}^{(\mu)}}U^{\mu\nu}(|{{\mbox{\boldmath$R$}}}_{i,\alpha}^{(\mu)}-{{\mbox{\boldmath$R$}}}_{j,\beta}^{(\nu)}|) \nonumber \\
&&-\sum_{i,\mu\in \alpha}  \dfrac{\partial}{\partial {{\mbox{\boldmath$R$}}}_{i,\alpha}^{(\mu)}}W(|{{\mbox{\boldmath$R$}}}_{i,\alpha}^{(\mu)}-{{\mbox{\boldmath$R$}}}_{p}|), 
\end{eqnarray}
\begin{eqnarray}
{{\mbox{\boldmath$N$}}}_{\alpha, {int}} &=& -\sum_{\beta\ne \alpha}\sum_{i,\mu\in \alpha}\sum_{j,\nu\in \beta} ({{\mbox{\boldmath$R$}}}_{i,\alpha}^{(\mu)} - {{\mbox{\boldmath$R$}}}_{\alpha}^{(G)}) \nonumber\\
&& \times \dfrac{\partial}{\partial {{\mbox{\boldmath$R$}}}_{i,\alpha}^{(\mu)}}U^{\mu\nu}(|{{\mbox{\boldmath$R$}}}_{i,\alpha}^{(\mu)}-{{\mbox{\boldmath$R$}}}_{j,\beta}^{(\nu)}|)\nonumber \\
&&-\sum_{i,\mu\in \alpha}({{\mbox{\boldmath$R$}}}_{i,\alpha}^{(\mu)} - {{\mbox{\boldmath$R$}}}_{\alpha}^{(G)})   \times\dfrac{\partial}{\partial {{\mbox{\boldmath$R$}}}_{i,\alpha}^{(\mu)}} W(|{{\mbox{\boldmath$R$}}}_{i,\alpha}^{(\mu)}-{{\mbox{\boldmath$R$}}}_{p}|), \nonumber\\
\end{eqnarray}

where $i,j=1,2,3$ and $\mu,\nu=b,f$. 
Here, $U^{\mu\nu}$ is the interaction potential between two spheres in which each comprise the body or the flagellum part of different swimmers, and $W$ is the interaction potential between such a sphere and the probe sphere. 
The explicit forms of $U^{\mu\nu}$ and $W$ are provided below. 
The active force acting on the body part, ${{\mbox{\boldmath$F$}}}_{\alpha, {A}}^{(b)}$, is given as 
\begin{eqnarray}
{{\mbox{\boldmath$F$}}}_{\alpha, {A}}^{(b)} = F_A {\hat {\mbox{\boldmath$n$}}}_{\alpha}.  \label{active_forceB}
\end{eqnarray} 
Equations (\ref{active_forceF}) and (\ref{active_forceB}) prescribe a force dipole $F_A\ell_0{\hat {\mbox{\boldmath$n$}}}_{\alpha}$ with $\ell_0{\hat {\mbox{\boldmath$n$}}}_{\alpha}={{\mbox{\boldmath$R$}}}_{\alpha}^{(G)}-{{\mbox{\boldmath$R$}}}_{\alpha}^{(CF)}$ [see also Eq. (\ref{active_stress})].  

Similarly, for the probe particle, the equations of motion for the center-of-mass velocity ${{\mbox{\boldmath$V$}}}_{p}$ and the angular velocity with respect to the center-of-mass ${{\mbox{\boldmath$\Omega$}}}_{p}$ are 
\begin{eqnarray}
M_p \dfrac{d{{\mbox{\boldmath$V$}}}_{p}}{dt} &=& {{\mbox{\boldmath$F$}}}_{{p,H}} + {{\mbox{\boldmath$F$}}}_{p,{int}} + F_{ex}{\hat{\mbox{\boldmath$x$}}},  
\label{VGp}  \\ 
I_p  \dfrac{d{{\mbox{\boldmath$\Omega$}}}_{p}}{dt} &=&  {{\mbox{\boldmath$N$}}}_{p, {H}},   \label{OGp}
\end{eqnarray}
where
\begin{eqnarray}
M_p = \rho{\mathcal V}_{p} \label{massp} 
\end{eqnarray} 
and 
\begin{eqnarray}
I_p =  \dfrac{2}{3}\int {\rm d}{{\mbox{\boldmath$r$}}} \rho\Psi_{p}({{\mbox{\boldmath$r$}}})|\Delta{{\mbox{\boldmath$r$}}}_p |^2 
\end{eqnarray}
are the mass and the moment of inertia of the probe particle, respectively. 
Here, $\Delta{{\mbox{\boldmath$r$}}}_p= {{\mbox{\boldmath$r$}}}-{{\mbox{\boldmath$R$}}}_p$, ${\mathcal V}_{p}=\int {\rm d}{{\mbox{\boldmath$r$}}}\Psi_p({{\mbox{\boldmath$r$}}})$ is the volume of the probe particle, and the probe particle density is also assumed to be the same as the solvent density. 
In Eq. (\ref{VGp}),  ${{\mbox{\boldmath$F$}}}_{p, {int}}$ is the force due to the interaction with swimmers: 
\begin{eqnarray}
{{\mbox{\boldmath$F$}}}_{p, {int}} = 
-\sum_{\alpha=1}^N\sum_{i,\mu\in \alpha}  \dfrac{\partial}{\partial {{\mbox{\boldmath$R$}}}_{p}}W(|{{\mbox{\boldmath$R$}}}_{i,\alpha}^{(\mu)}-{{\mbox{\boldmath$R$}}}_{p}|),  
\end{eqnarray}
where $i=1,2,3$ and $\mu=b,f$.

We assume the following form of the interparticle potentials: 
\begin{eqnarray}
U^{\mu\nu}(r) = \epsilon(1-\delta_{\mu,f}\delta_{\nu,f}) \biggl(\dfrac{2a}{r}\biggr)^{12},  \label{potentialU} 
\end{eqnarray}
where $\epsilon$ is a positive energy constant, $\delta_{\mu,f}$ is the Kronecker delta, and $\mu,\nu=b,f$. 
This form prevents the body part of a swimmer from overlapping on different swimmers but allows overlaps among the flagellum parts. 
The interaction potential $W$ between the probe particle and the particles constituting a swimmer is introduced to prevent the penetration of swimmers through the probe boundary. In this study, $W$ is assumed to be given as 
\begin{eqnarray}
W(r) &=& \epsilon \biggl[\dfrac{2a}{r-(R-a)}\biggr]^{12},  \label{potentialW}
\end{eqnarray}
where we assume the same energy constant as that of $U^{\mu\nu}$. 

For a direct comparison between simulations with and without HIs, we have made equivalent simulations without HIs, using the same parameters for the interactions. 
Furthermore, in Eqs. (\ref{VG}), (\ref{OG}), (\ref{VGp}), and (\ref{OGp}), the hydrodynamic forces and torques are replaced as 
\begin{eqnarray}
  {{\mbox{\boldmath$F$}}}_{\alpha, {H}} &\rightarrow& -\zeta_{||}{\hat {\mbox{\boldmath$n$}}}_{\alpha}{\hat {\mbox{\boldmath$n$}}}_{\alpha}\cdot {{\mbox{\boldmath$V$}}}_\alpha^{(G)} - \zeta_{\bot}({\stackrel{\leftrightarrow}{\mbox{\boldmath$\delta$}}}- {\hat {\mbox{\boldmath$n$}}}_{\alpha}{\hat {\mbox{\boldmath$n$}}}_{\alpha})\cdot {{\mbox{\boldmath$V$}}}_\alpha^{(G)},  
\nonumber \\ \label{woVG} \\
 {{\mbox{\boldmath$N$}}}_{\alpha, {H}} &\rightarrow& -\zeta_R{{\mbox{\boldmath$\Omega$}}}_{\alpha}^{(G)},   \label{woOG} \\
 {{\mbox{\boldmath$F$}}}_{p, {H}} &\rightarrow& -6\pi \eta_s R  {{\mbox{\boldmath$V$}}}_p,  \label{woVP} \\ 
 {{\mbox{\boldmath$N$}}}_{p, {H}} &\rightarrow& -8\pi \eta_s R^3 {{\mbox{\boldmath$\Omega$}}}_p.    \label{woOP}
\end{eqnarray}
Here, the values of the friction coefficients $\zeta_{||}$, $\zeta_\bot$, and $\zeta_R$ are numerically evaluated as those of an isolated swimmer with HIs;  
$\zeta_{||}=9.6\pi\eta_s a $, $\zeta_{\bot}=10.9\pi\eta_s a$, and $\zeta_R=38.7\pi\eta_s a^3$. 

In our simulations, as illustrated in Fig. 1(c), periodic boundary conditions are imposed in all directions with the linear dimension $L$.  We make the equations dimensionless by measuring space and time in units of $h$, which is the discretization mesh size used when solving Eqs. (\ref{Navier_Stokes})-(\ref{incompressibility}), and $t_0=\rho h^2/\eta_s$, which is the momentum diffusion time across the unit length, respectively. Accordingly, the scaled solvent viscosity is $1$, and the units of velocity, stress, force, and energy are chosen to be $h/t_0$, $\rho h^2/t_0^2$, $\rho h^4/t_0^2$ and $\rho h^5/ t_0^2$, respectively. 
In our simulations, we set $\epsilon=30$ and $F_A=20$. The parameters determining the swimmer's shape are set to be $a=3.2$, $\ell_0=| {{\mbox{\boldmath$R$}}}_\alpha^{(G)}- {{\mbox{\boldmath$R$}}}_\alpha^{(CF)}|=4.5a$ and $\xi=0.5$. 
In this study, the swimmer volume fraction is identified as that of the rigid body particles given by $\phi=\sum_{\alpha=1}^N{\mathcal V}_\alpha^{(b)}/(L^3-4\pi R^3/3)=N{\mathcal V}_\alpha^{(b)}/(L^3-4\pi R^3/3)$.

\section{Results}
\subsection{Friction coefficient}
\begin{figure*}[hbt] 
\centering
\includegraphics[width=17cm]{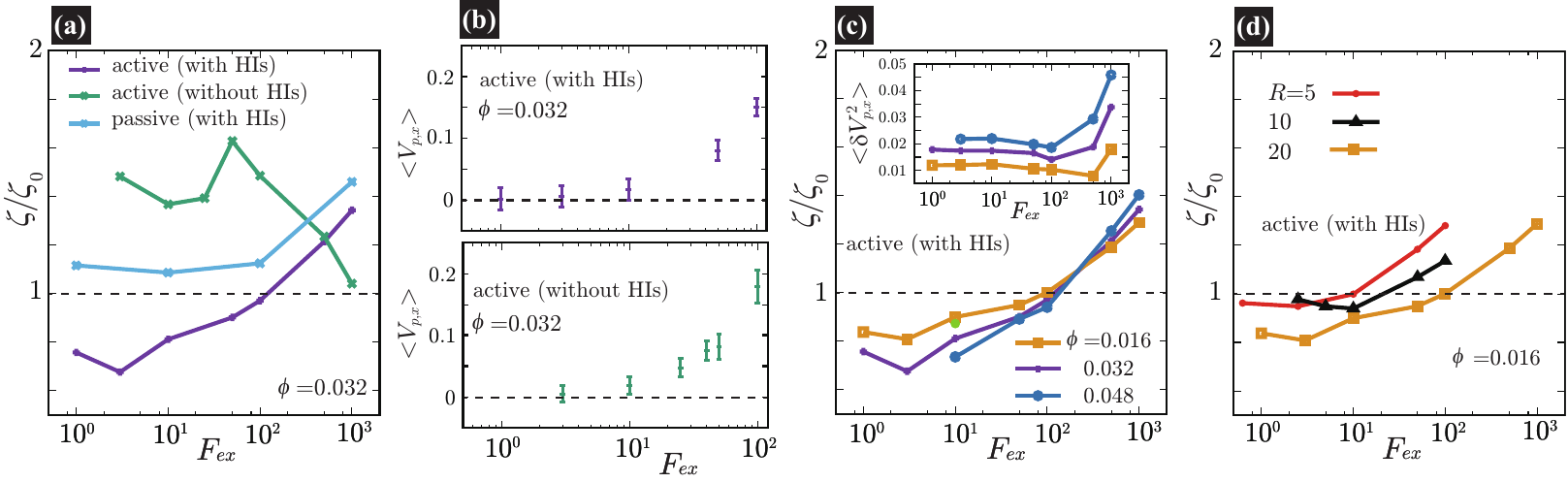}
\caption{In (a), (c), and (d), we present the friction coefficients $\zeta$ scaled by $\zeta_0$ against $F_{ex}$. For the cases with HIs, $\zeta_0$ is the friction coefficient of the probe particle suspended in a pure solvent with $F_{ex}=10$. On the contrary, in the cases without HIs, $\zeta_0=6\pi\eta_s R$ as set in Eq. (\ref{woVP}).  
In (a), $\zeta/\zeta_0$ is shown for three different conditions at a probe radius $R=20(=6.25a)$ and $\phi=0.032$.  
With HIs, $\zeta$ is significantly smaller than $\zeta_0$ for $F_{ex}\lesssim 10^2$, contrasting with the behavior in the absence of HIs. 
We also present a passive case, where we use the same swimmer model but without the active force ($F_A=0$). 
In (b), we plot $\langle V_{p,x}\rangle$, with error bars, for $F_{ex}\le 10^2$ both with and without HIs, shown in the upper and lower panels, respectively. For relatively smaller $F_{ex}$, strong fluctuations in $V_{p,x}$ are found, reflecting significant back-and-forth motions due to frequent collisions with surrounding swimmers. 
However, each data point is derived from simulations conducted over extended periods to ensure the accuracy of $\langle V_{p,x}\rangle$; especially for smaller $F_{ex}$ values like $F_{ex}=3$ and 10, the simulation ran for a period where the probe particle traveled several hundred times its own size. 
Furthermore, it is noteworthy that, in passive ($F_A=0$) suspensions, fluctuations in $V_{p,x}$ (not shown here) are significantly smaller than those in active suspensions within the same $F_{ex}$ range. 
In (c),  $\zeta/\zeta_0$ is shown for different $\phi$ at a probe radius $R=20(=6.25a)$ with HIs.  
The main data are the results for $L=128$, while the yellow-green closed circle represents the data for a larger system size $L=256$ at $R=20$, $\phi=0.016$, and $F_{ex}=10$, almost corresponding to the case of $L=128$ at the same $R$ and $\phi$. The inset displays $\langle \delta V_{p,x}^2\rangle$ with $\delta V_{p,x}=  V_{p,x}-\langle V_{p,x}\rangle$. For relatively small $F_{ex}$, where the average probe velocity $\langle V_{p,x}\rangle$ is considerably smaller than the average speed of microswimmers ($\sim 0.1$), $\langle \delta V_{p,x}^2\rangle$  is nearly constant and smaller for lower $\phi$. This feature suggests that the fluctuations in probe velocities are determined by the details of probe-swimmer collisions and their statistics.   
 (d) $\zeta/\zeta_0$ versus $F_{ex}$ is shown for several different probe sizes ($R=5$, 10, and 20) at $\phi=0.016$. At $R=5$ and $10$, within the examined range of $F_{ex}$, the reduction of $\zeta$ is less pronounced than at $R=20$.}
\label{Fig2}
\end{figure*}

Figures 2(a), (c), and (d) illustrate the scaled friction coefficient $\zeta/\zeta_0$ under various conditions. In this study, the friction coefficient of the probe particle $\zeta$ is defined as 
\begin{eqnarray}
\zeta=\dfrac{F_{ex}}{\langle V_{p,x} \rangle}, 
\end{eqnarray} 
where $V_{p,x}$ is the $x$-component of the probe velocity and $\langle \cdots \rangle$ represents the time average in a steady state. 
Figure 2(b) shows the average probe velocity $\langle V_{p,x}\rangle$ at a probe radius $R=20(=6.25a)$ and $\phi=0.032$, both with and without HIs. 
In the cases with HIs, $\zeta_0$ represents the bare friction coefficient experienced by the probe particle when suspended in a pure solvent. Notably, for a relatively small $F_{ex}$, $\zeta$ is significantly lower than $\zeta_0$. As shown in Fig. 2(c), this reduction is enhanced as the swimmer volume fraction $\phi$ increases at least within the range examined in our simulations ($\phi\lesssim 0.05$). 
On the contrary, in the cases without HIs, $\zeta$ is always larger than $\zeta_0$, showing non-monotonic $F_{ex}$ dependence \cite{comment_zeta_woHIs} similar to that observed in Ref. \cite{Burkholder2}.  
The observed distinction strongly suggests that HIs play a significant role in the reduction of $\zeta$ in active suspensions. 

In macrorheology simulations using the same rod-like model swimmer, HIs induce a weak alignment of swimmer orientations along the elongation axis of the applied flow field, resulting in the acceleration of the mean flow and a subsequent reduction in the viscosity \cite{Hayano_Furukawa}. Accordingly, we can expect that the decrease in $\zeta$ in the present microrheology simulations also reflects a similar mechanism involving some ordering of swimming directions in the bulk region. However, as shown below, this is not the case. 

Figure 2(d) shows $\zeta/\zeta_0$ against $F_{ex}$ for three different probe sizes.  
For the $F_{ex}$ ranges at $R=5$ and $10$, the magnitude of the induced velocity gradient in the bulk region (more than one swimmer size away from the probe surface) approximately corresponds to that observed for $F_{ex}\sim 10-10^3$ in Figs. 2(a) and (c), but the reduction in $\zeta$ is not as noticeable as for $R=20$. 
Furthermore, in Fig. 2(c), the main data are the results for $L=128$, while the yellow-green closed circle represents the data for a larger system size $L=256$ at $R=20$, $\phi=0.016$, and $F_{ex}=10$. Although we examine only a single case, a significant system-size dependence is hardly observed, which contrasts with the macrorheology case, where the viscosity reduction shows a strong system size dependence \cite{Hayano_Furukawa}. 
These findings suggest that the drag-reduction mechanism depends on the local probing details and thus qualitatively differs from that of the viscosity reduction observed through macrorheology measurements.

\begin{figure} 
\includegraphics[width=9cm]{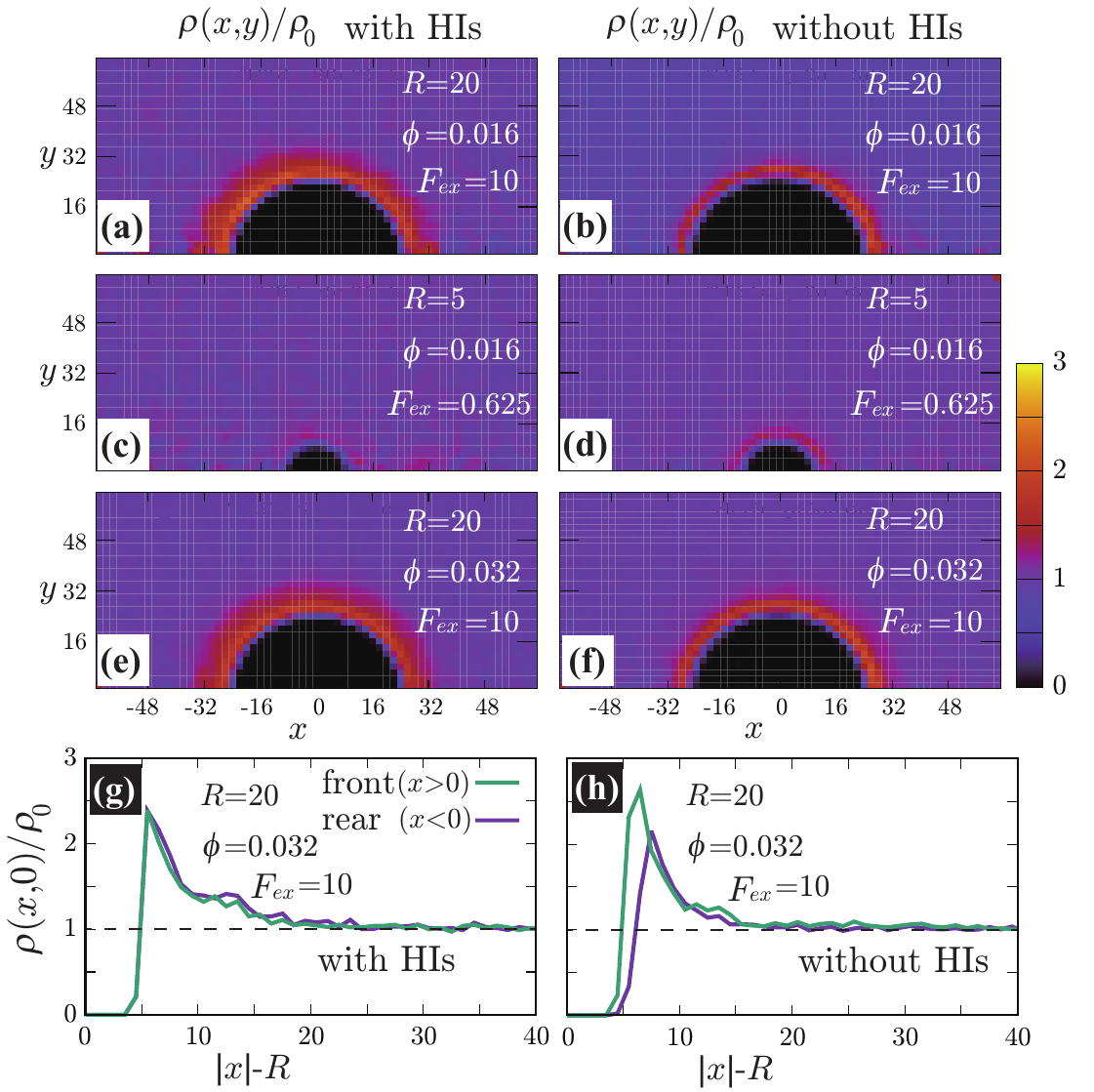}
\caption{(a)-(f) Contour plots of the scaled number density $\rho(x,y)/\rho_0$ for various conditions, where $\rho_0$ is the average density defined as $\rho_0=N/[L^3-4\pi R^3/3]$. The left and right panels correspond to cases with and without HIs, respectively. We assume axial symmetry about the $x$-axis.  
At $R=20(\gtrsim \ell_0$), as shown in (a), (b), (e), and (f), $\rho(x,y)$ near the probe boundary is considerably larger than $\rho_0$, irrespective of the presence of HIs. On the other hand, for $R=5(\lesssim \ell_0)$, as shown in (c) and (d), the peak of $\rho(x,y)$ near the surface is less pronounced. In (g) and (h), $\rho(x,0)/\rho_0$ is shown at $R=20$, $\phi=0.032$, and $F_{ex}=10$ with and without HIs, respectively.}
\label{Fig3}
\end{figure}

Then, to understand how $\zeta$ is reduced, let us investigate the local swimmer states around the probe and their impact on the probe dynamics. In Figs. 3(a)-(f), we show the contour plots of the number density $\rho(x,y)$. Hereafter, we assume that axial symmetry about the $x$-axis approximately holds. At $R=20(\gtrsim \ell_0)$, as shown in Figs. 3 (a), (b), (e), and (f), $\rho(x,y)$ near the probe boundary is considerably larger than in the bulk region, irrespective of the presence of HIs. As indicated in the literature, both the steric \cite{Ji-Tang} and hydrodynamic \cite{Berke} effects cause the entrapment of microswimmers on the boundaries. A self-propulsive rod-like particle, upon colliding with the probe, cannot rebound immediately due to the steric/geometric constraints. Instead, it stays near the probe for a while, swimming along the surface. This behavior may result in a larger density near the probe surface. Note that HIs between the probe and the swimmers enhance the entrapment \cite{Berke}, but they are not significantly dominant in the present condition. Also, as shown in Figs. 3 (g) and (h), without HIs, the increase in $\rho(x,y)$ is slightly more pronounced on the front than the rear, but such a behavior diminishes with HIs. At $R=5(\lesssim \ell_0)$, as shown in Figs. 3(c) and (d), the peak of $\rho(x,y)$ near the boundary decreases, both with and without HIs, suggesting a weaker entrapment effect for smaller $R$.

\subsection{Local swimmer states arond the probe particle}

\begin{figure}[hbt] 
\includegraphics[width=8cm]{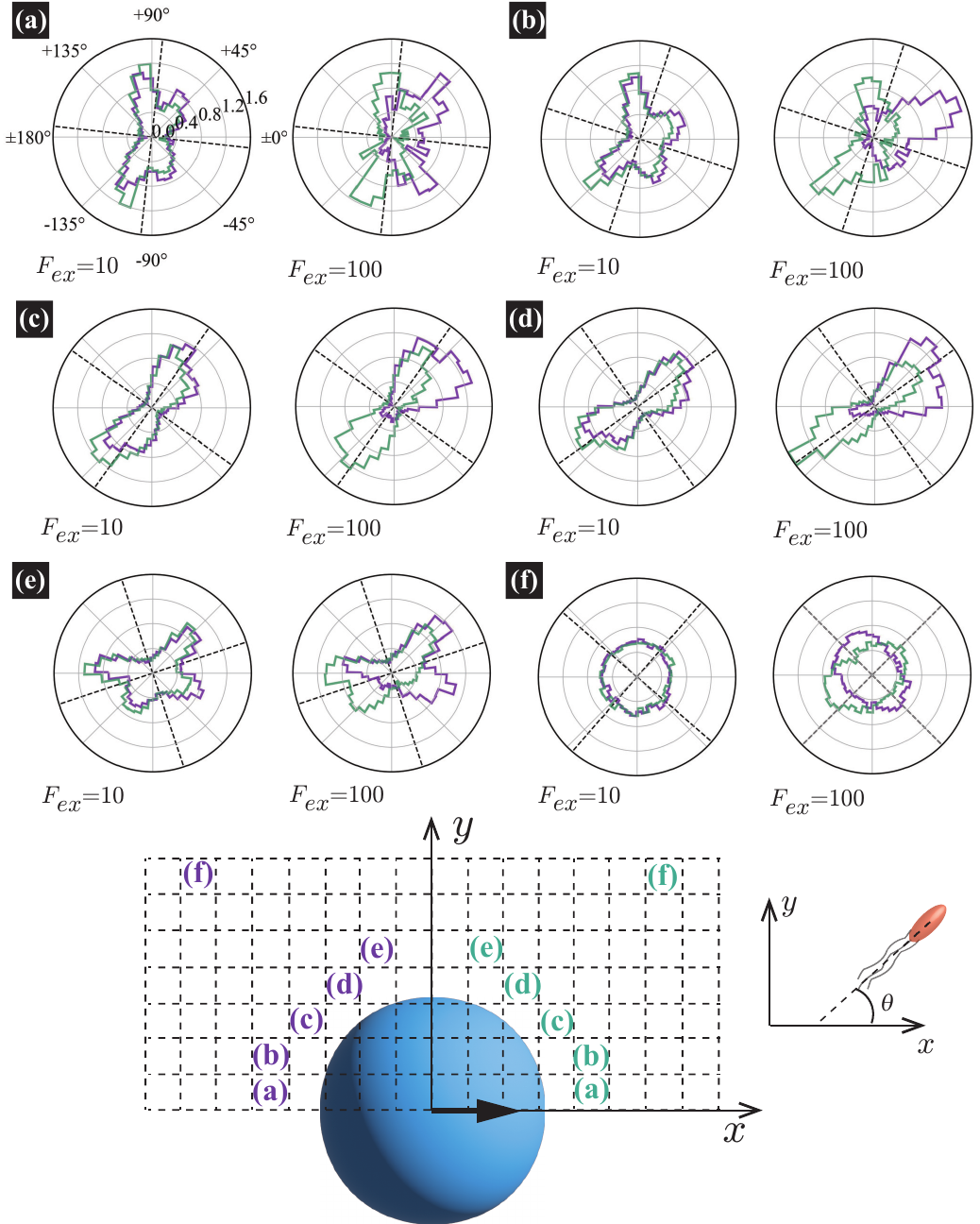}
\caption{The orientation distributions of swimmers $P(\theta;x,y)$ averaged over a square region outlined by dashed lines at $R=20$ and $\phi=0.032$ with HIs, with the block size being $2a=0.32R(=6.4)$. For the definition of $P(\theta;x,y)$, please refer to the explanation presented in the main text. The probe-sphere's center is located at $(x,y)=(0,0)$. In (a)-(f), the left and right panels present the distributions at $F_{ex}=$10 and $100$, respectively. The violet and dark-green lines correspond to $P(\theta;x,y)$ $(x<0)$ and $P(180^\circ-\theta;x,y)$ $(x>0)$, respectively, calculated in the regions indicated by identical colored characters and normalized to make the total enclosed area equal $1$. The dotted lines guide the normal and tangential directions along the probe sphere.
}
\label{Fig4}
\end{figure}

Figure 4 shows the orientation distribution of swimmers $P(\theta;x,y)$ around the probe particle. It is defined as $P(\theta;x,y)=c(x,y)\sum_{\alpha}\langle \delta[\theta-\cos^{-1}({\hat {\mbox{\boldmath$t$}}}_\alpha\cdot {\hat {\mbox{\boldmath$x$}}})]\delta({\mbox{\boldmath$r$}}-{\mbox{\boldmath$R$}}_\alpha)\rangle$, 
where ${ {\mbox{\boldmath$r$}}}=(x,y,0)$, ${\hat {\mbox{\boldmath$x$}}}$ is the unit vector along the $x$-axis and ${\hat {\mbox{\boldmath$t$}}}_\alpha={\hat {\mbox{\boldmath$n$}}}_\alpha^{||}/|{\hat {\mbox{\boldmath$n$}}}_\alpha^{||}|$ with ${\hat {\mbox{\boldmath$n$}}}_\alpha^{||}$ denoting the projected swimming direction onto the $x$-$y$ plane.  
The normalization factor $c(x,y)$ specific to the present visualization in Fig. 4 is determined so that $(1/2)\int {\rm d}\theta P^2=1$. 
The main text presents the results for two cases at $F_{ex}=10$ and $100$ with HIs. For cases without HIs, please refer to the Appendix C. 
In Fig. 4, the panels (a)-(e) show $P(\theta;x,y)$ near the probe boundary.
The tilt angle of the swimmers (measured with respect to the tangential direction of the probe surface) is larger when they face the probe than when they face the bulk, but this distinction is subtle in (c) and (d) for $F_{ex}=10$. Additionally, more swimmers face the probe for $x<0$ than $x>0$.
These characteristics are more evident at $F_{ex}=100$ than at $F_{ex}=10$.
The (repulsive) HIs are more intensified when a swimmer approaches the probe particle from the front rather than the rear. In this situation, swimmers suffer from significant scatterings at the front side, while scatterings are comparatively weaker at the rear. This difference in scattering causes the observed front-rear asymmetry in $P(\theta;x,y)$. Supporting simulations are provided in the Appendix B. 

\subsection{Driving force exerted on the probe particle by swimmers}

The observed local swimmers' behavior significantly affects the probe motions. 
To illustrate this, we display the contour plots of the $rr$- and $r\theta$-components of the active stresses, denoted as $s^A_{rr}(x,y)$ and $s^A_{r\theta}(x,y)$, respectively, for $F_{ex}=10$ and $\phi=0.032$ in Fig. 5. 

We here derive an expression for the local active stress tensor defined in a small subsystem denoted as $K$, with a volume $V_K$ and a representative position of $(x,y,z)$. 
Referring to a similar procedure given in Refs. \cite{SPM2,Hayano_Furukawa}, 
the active stress can be defined as 
\begin{eqnarray}
{\stackrel{\leftrightarrow}{\mbox{\boldmath$s$}}_A} &=& -\dfrac{1}{V_K} \sum_{\alpha\in K}  F_A{\hat {\mbox{\boldmath$n$}}}_\alpha  \int_{K} {\rm d}{{\mbox{\boldmath$r$}}} {{\mbox{\boldmath$r$}}}  \biggl[\dfrac{\Psi^{(b)}_\alpha ({{\mbox{\boldmath$r$}}})}{{\mathcal V}_\alpha^{(b)}} - \dfrac{\Psi^{(f)}_\alpha ({{\mbox{\boldmath$r$}}})}{{\mathcal V}_\alpha^{(f)}}\biggr] \nonumber \\
&=& -\dfrac{1}{V_K} \sum_{\alpha\in K}  F_A{\hat {\mbox{\boldmath$n$}}}_\alpha  \biggl({{\mbox{\boldmath$R$}}}_{\alpha}^{(G)} - {{\mbox{\boldmath$R$}}}_{\alpha}^{(CF)}\biggr) \nonumber \\
&=& -\dfrac{1}{V_K} \sum_{\alpha\in K}  F_A\ell_0 
{\hat {\mbox{\boldmath$n$}}}_\alpha {\hat {\mbox{\boldmath$n$}}}_\alpha, \label{active_stress}
\end{eqnarray}
where $\alpha\in K$ indicates that the position ${{\mbox{\boldmath$R$}}}_{\alpha}$ is within the region $K$. 
By using the conversion formula to transform  the tensor from Cartesian to polar coordinates, we obtain the following expressions 
\begin{eqnarray}
\langle s_{A,rr}\rangle &=&  \dfrac{1}{2}(\langle s_{A,xx}\rangle+\langle s_{A,yy}\rangle) \nonumber \\&&+\dfrac{1}{2} (\langle s_{A,xx}\rangle-\langle s_{A,yy}\rangle)\cos 2\theta + \langle s_{A,xy}\rangle \sin 2\theta, \nonumber \\\\
\langle s_{A,r\theta}\rangle &=&  \langle s_{A,xy}\rangle \cos 2\theta
-\dfrac{1}{2}(\langle s_{A,xx}\rangle -  \langle s_{A,yy}\rangle) \sin 2\theta. \nonumber\\
\end{eqnarray}
Here, we assume axial symmetry about the $x$-axis of the local swimmers properties in steady states.  
In our calculations, we evaluate the active stress in a small region with a linear dimension of $a$, which is not large enough to define an instantaneous stress tensor. However, this does not pose any practical problem when investigating the average properties of the active stress in steady states.   
\begin{figure}[tb] 
\centering
\includegraphics[width=8cm]{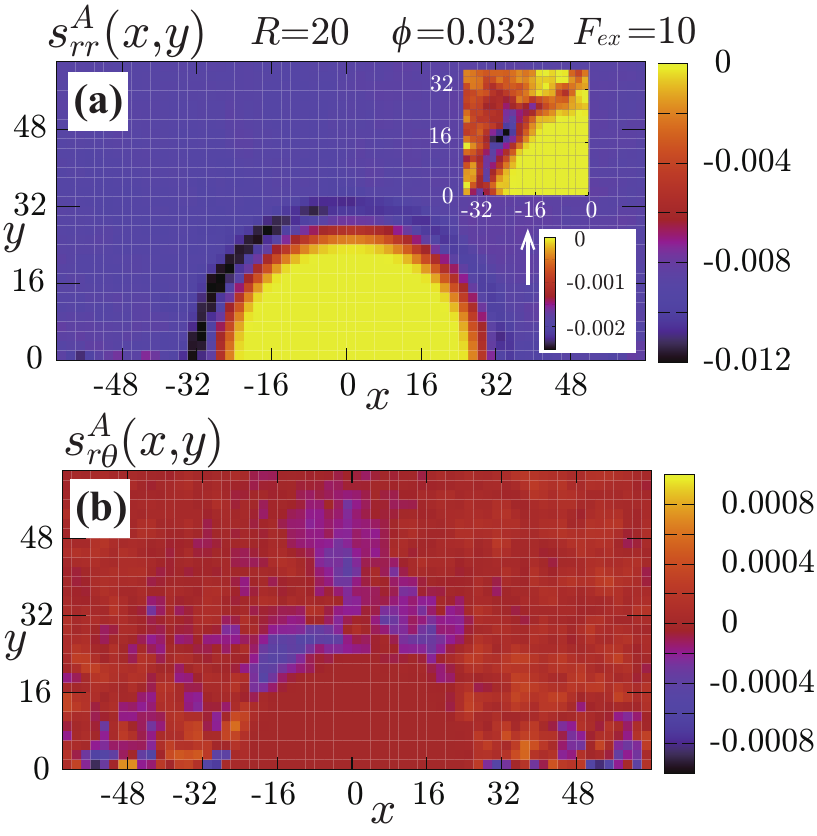}
\caption{Contour plots of the $rr$- and $r\theta$-components of the active stresses, denoted as $s^A_{rr}(x,y)$ (a) and $s^A_{r\theta}(x,y)$ (b), respectively, for $F_{ex}=10$ and $\phi=0.032$. 
In (a), the inset shows $s^A_{rr}(x,y)-s^A_{rr}(-x,y)$, indicating apparent front rear asymmetry in $s^A_{rr}(x,y)$. This asymmetry means that the swimmers exert more force from the rear. 
In (b), $s^A_{r\theta}(x,y)$ shows more significant fluctuations than $s^A_{rr}(x,y)$. 
}
\label{Fig5}
\end{figure}

The $rr$-component $s^A_{rr}(x,y)$ reflects the degree of the extent to which the swimmers face the probe sphere, either from the tail or the head. As shown in Fig. 5(a), around the probe, the magnitude of $s^A_{rr}(x,y)$ is slightly less pronounced at the front side than at the rear, reflecting the front rear asymmetry in $P(\theta;x,y)$. 
This behavior intuitively suggests that the swimmers exert more force on the probe from the rear than from the front, which is more clearly seen in the inset plotting $s^A_{rr}(x,y)-s^A_{rr}(-x,y)$. 
On the other hand, as shown in Fig. 5(b), $s^A_{r\theta}(x,y)$ exhibits a notable negative value around the regions indicated as (c)-(e) for $x<0$ in Fig. 4.
Note, however, that despite averaging over a period in which the probe particle moves several hundred times more than $R$, $s^A_{r\theta}(x,y)$ still exhibits significant fluctuations. 
The contribution from the active stress to the driving force is estimated by:
\begin{eqnarray}
\Delta F\sim \int_{S_p} {\rm d}S ( s^A_{rr}\cos\theta -s^A_{r\theta} \sin\theta), \label{driving}
\end{eqnarray}
where $S_p$ represents the surface at which the peak of $\rho(x,y)$ appears and ${\rm d}S$ is the surface area differential on $S_p$. This $\Delta F$ is roughly identified with an additional driving force to $F_{ex}$.
The magnitude of $s^A_{r\theta}(x,y)$ is approximately 0.1 of $s^A_{rr}(x,y)$ on $S_p$, and $\Delta F$ mostly depends on $s^A_{rr}(x,y)$.
According to the inset of Fig. 5(a), the contribution from $s^A_{rr}(x,y)$ is estimated to be on the order of unity, using $s^A_{rr}(x,y)-s^A_{rr}(-x,y)\sim 10^{-3}$ on $S_p$. This estimation explains the notable reduction in $\zeta$ at $F_{ex}=10$.
As $F_{ex}$ increases, the front rear asymmetry in the active stress is enhanced, but $\Delta F$ may not linearly increase in $F_{ex}$, resulting in the less noticeable contributions to the reduction in $\zeta$.

\section{Concluding Remarks}

Let us evaluate the applicability of our results to realistic situations by considering a  typical experimental setup of dilute {\it E. coli} suspensions with a volume fraction similar to our simulations ($\sim 0.01$). 
Here, we assume that the swimming speed of {\it E. coli} is $v_s\sim10$ $\mu$m/s, the magnitude of the active force is $F_A\sim$1 pN, and the cell size is 
$\ell_0\sim$ 1 $\mu$m. In a typical condition of magnetic-bead microrheology, the force exerted on the probe is in the pN range, and magnetic colloidal particles have diameters ranging from $1$ to $10$ $\mu$m ($=2R$). In our simulation conditions, where the drag reduction is observed, the force ratio is $F_{ex}/F_A\lesssim 3$ and the size ratio is $2R/\ell_0\sim 3$, and these can be met in real experiments \cite{MicrorheologyB}. 
Therefore, we can expect that a substantial drag reduction occurs, if similar density and orientational distributions to those obtained in our simulations are realized in practical experiments.

In summary, using a pusher-type swimmer model, the present study investigated the microrheology of active suspensions through direct hydrodynamic simulations. We revealed that, with HIs, the friction coefficient of the probe particle can be significantly smaller than when immersed in a pure solvent for relatively small drag forces. The local swimmer states near the probe boundary predominantly influence the observed drag reduction. That is, in active suspensions, microrheology measurements are strongly influenced by local probing details and thus can be qualitatively different from macrorheology measurements.
Hydrodynamic interactions induce the front rear asymmetry in the swimmer  orientation distributions. This asymmetry further causes a force/stress imbalance, resulting in an additional driving force acting on the probe particle through solvent-mediated interactions (HIs). However, without HIs, the local states of swimmers outside the (short) range of the direct interaction potential do not contribute to the friction force. Our simulation studies illuminate the fundamental importance of HIs in active suspensions and suggest potential microfluidics applications. 
These issues will be further explored in subsequent studies.

\section*{Acknowledgements}

This work was supported by KAKENHI (Grants No. 20H05619) from the
Japan Society for the Promotion of Science (JSPS). 
We are also grateful for UROP (Undergraduate Research Opportunity Program) of the Institute of Industrial Science, the University of Tokyo. 

\appendix

\section{Explicit time-integration algorithm}
\begin{widetext}
Here, following Refs. 22, 47, and 48,  we describe an explicit time-integration scheme for solving model equations as follows. 
The set of physical variables is assumed to be clearly defined at the discrete time step $t_n=n \Delta t$. 

First, we solve Eq. (5) in the main text without including ${{\mbox{\boldmath$f$}}}_{H}$ as 
\begin{eqnarray}
{{\mbox{\boldmath$v$}}}^* = {{\mbox{\boldmath$v$}}}_n 
+ \dfrac{1}{\rho}\int _{t_n}^{t_{n+1}} {\rm d}s \biggl(- \rho{{\mbox{\boldmath$v$}}} \cdot {{\mbox{\boldmath$\nabla$}}} {{\mbox{\boldmath$v$}}} +  {{\mbox{\boldmath$\nabla$}}}\cdot {\stackrel{\leftrightarrow}{\mbox{\boldmath$\Sigma$}}_{vis}} + {{\mbox{\boldmath$f$}}}_{A}^{(f)} \biggr)^{\bot},  \nonumber \\
\end{eqnarray}
where ${{\mbox{\boldmath$v$}}}_n$ is the velocity field at $t=t_n$ and $(\cdots)^\bot$ denotes taking the transverse part. 

 Second, we update ${{\mbox{\boldmath$R$}}}_\alpha^{(G)}$, ${\hat {\mbox{\boldmath$n$}}}_\alpha$, and ${{\mbox{\boldmath$R$}}}_p$ as 
\begin{eqnarray}
{{\mbox{\boldmath$R$}}}_\alpha^{(G)}(t_{n+1}) &=& {{\mbox{\boldmath$R$}}}_\alpha^{(G)}(t_{n}) +  \int _{t_n}^{t_{n+1}} {\rm d}s {{\mbox{\boldmath$V$}}}_\alpha^{(G)}, \\
{\hat {\mbox{\boldmath$n$}}}_\alpha(t_{n+1}) &=& {\hat {\mbox{\boldmath$n$}}}_\alpha(t_n) +  \int _{t_n}^{t_{n+1}} {\rm d}s  {{\mbox{\boldmath$\Omega$}}}_\alpha^{(G)} \times  {\hat {\mbox{\boldmath$n$}}}_\alpha, \\ 
{{\mbox{\boldmath$R$}}}_p(t_{n+1}) &=& {{\mbox{\boldmath$R$}}}_p(t_{n}) +  \int _{t_n}^{t_{n+1}} {\rm d}s {{\mbox{\boldmath$V$}}}_p. 
\end{eqnarray}
With these updated ${{\mbox{\boldmath$R$}}}_\alpha^{(G)}$, ${\hat {\mbox{\boldmath$n$}}}_\alpha$, and ${{\mbox{\boldmath$R$}}}_p$, 
we also update $\Psi_\alpha^{(b)}({{\mbox{\boldmath$r$}}})$, $\Psi_\alpha^{(f)}({{\mbox{\boldmath$r$}}})$, and $\Psi_p({{\mbox{\boldmath$r$}}})$.  

 Third, the particle velocities and angular velocities are updated by solving Eqs. (9), (10), (16), and (17) in the main text as 
 
\begin{eqnarray}
{{\mbox{\boldmath$V$}}}_\alpha^{(G)}(t_{n+1}) &=& {{\mbox{\boldmath$V$}}}_\alpha^{(G)}(t_{n})+\dfrac{1}{M_\alpha}\int _{t_n}^{t_{n+1}} {\rm d}s ({{\mbox{\boldmath$F$}}}_{\alpha, {H}} + {{\mbox{\boldmath$F$}}}_{\alpha, {int}}+ {{\mbox{\boldmath$F$}}}_{\alpha, {A}}^{(b)}) , \label{VG2} \\
{{\mbox{\boldmath$\Omega$}}}_\alpha^{(G)}(t_{n+1})&=&{{\mbox{\boldmath$\Omega$}}}_\alpha^{(G)}(t_{n})+{\stackrel{\leftrightarrow}{\mbox{\boldmath$I$}}^{-1}_{\alpha}}\cdot 
 \biggl[ \int _{t_n}^{t_{n+1}} {\rm d}s ({{\mbox{\boldmath$N$}}}_{\alpha, {H}} + {{\mbox{\boldmath$N$}}}_{\alpha, {int}})\biggr],   \label{OG2} \\
 {{\mbox{\boldmath$V$}}}_p(t_{n+1}) &=& {{\mbox{\boldmath$V$}}}_p(t_{n})+\dfrac{1}{M_p}\int _{t_n}^{t_{n+1}} {\rm d}s ({{\mbox{\boldmath$F$}}}_{p, {H}} + {{\mbox{\boldmath$F$}}}_{p, {int}}+ {{\mbox{\boldmath$F$}}}_{p, {ex}}) , \label{VP2} \\
{{\mbox{\boldmath$\Omega$}}}_p(t_{n+1})&=&{{\mbox{\boldmath$\Omega$}}}_p(t_{n})+\dfrac{1}{I_p}
  \int _{t_n}^{t_{n+1}} {\rm d}s {{\mbox{\boldmath$N$}}}_{p, {H}}.   \label{OP2}
\end{eqnarray} 
Here, the explicit forms of $\int _{t_n}^{t_{n+1}} {\rm d}s {{\mbox{\boldmath$F$}}}_{\alpha, {H}}$, $\int _{t_n}^{t_{n+1}} {\rm d}s {{\mbox{\boldmath$N$}}}_{\alpha, {H}}$, $\int _{t_n}^{t_{n+1}} {\rm d}s {{\mbox{\boldmath$F$}}}_{p, {H}}$ and $\int _{t_n}^{t_{n+1}} {\rm d}s {{\mbox{\boldmath$N$}}}_{p, {H}}$ are given as 

\begin{eqnarray}
\int _{t_n}^{t_{n+1}} {\rm d}s {{\mbox{\boldmath$F$}}}_{\alpha, {H}} = \int {\rm d}{{\mbox{\boldmath$r$}}} \rho \Psi_{\alpha,n+1}^{(b)} \biggl\{{{\mbox{\boldmath$v$}}}^* -\biggl[{{\mbox{\boldmath$V$}}}_\alpha^{(G)}(t_n)+ {{\mbox{\boldmath$\Omega$}}}_\alpha^{(G)}(t_n)\times \bigl( {{\mbox{\boldmath$r$}}}- {{\mbox{\boldmath$R$}}}_\alpha^{(G)}(t_{n+1})\bigr)  \biggr]  \biggr\}, \label{Hforce} 
\end{eqnarray}
\begin{eqnarray}
\int _{t_n}^{t_{n+1}} {\rm d}s {{\mbox{\boldmath$N$}}}_{\alpha, {H}} = \int {\rm d}{{\mbox{\boldmath$r$}}} \rho\Psi_{\alpha,n+1}^{(b)}  \bigl( {{\mbox{\boldmath$r$}}}- {{\mbox{\boldmath$R$}}}_\alpha^{(G)}(t_{n+1})\bigr) \times \biggl\{{{\mbox{\boldmath$v$}}}^* -\biggl[{{\mbox{\boldmath$V$}}}_\alpha^{(G)}(t_n)+ {{\mbox{\boldmath$\Omega$}}}_\alpha^{(G)}(t_n)\times  \bigl( {{\mbox{\boldmath$r$}}}- {{\mbox{\boldmath$R$}}}_\alpha^{(G)}(t_{n+1})\bigr) \biggr]  \biggr\},  \label{Htorque}
\end{eqnarray} 
\begin{eqnarray}
\int _{t_n}^{t_{n+1}} {\rm d}s {{\mbox{\boldmath$F$}}}_{p, {H}} = \int {\rm d}{{\mbox{\boldmath$r$}}} \rho \Psi_{p,n+1} \biggl\{{{\mbox{\boldmath$v$}}}^* -\biggl[{{\mbox{\boldmath$V$}}}_p(t_n)+ {{\mbox{\boldmath$\Omega$}}}_p(t_n)\times \bigl( {{\mbox{\boldmath$r$}}}- {{\mbox{\boldmath$R$}}}_p(t_{n+1})\bigr)  \biggr]  \biggr\},  \label{HforceP} 
\end{eqnarray}
and
\begin{eqnarray}
\int _{t_n}^{t_{n+1}} {\rm d}s {{\mbox{\boldmath$N$}}}_{p, {H}} = \int {\rm d}{{\mbox{\boldmath$r$}}} \rho\Psi_{p,n+1}  \bigl( {{\mbox{\boldmath$r$}}}- {{\mbox{\boldmath$R$}}}_p(t_{n+1})\bigr) \times \biggl\{{{\mbox{\boldmath$v$}}}^* -\biggl[{{\mbox{\boldmath$V$}}}_p(t_n)+ {{\mbox{\boldmath$\Omega$}}}_p(t_n)\times  \bigl( {{\mbox{\boldmath$r$}}}- {{\mbox{\boldmath$R$}}}_p(t_{n+1})\bigr) \biggr]  \biggr\},  \label{HtorqueP}
\end{eqnarray} 
where $\Psi_{\alpha,n+1}^{(b)}$ and $\Psi_{p,n+1}$ denote $\Psi_\alpha^{(b)}({{\mbox{\boldmath$r$}}})$ and $\Psi_p({{\mbox{\boldmath$r$}}})$ at $t=t_{n+1}$, respectively.   

Finally, we update the velocity field by embedding the rigid body motions in ${{\mbox{\boldmath$v$}}}^*$ through the body force ${{\mbox{\boldmath$f$}}}_{H}$ as 
\begin{eqnarray}
{{\mbox{\boldmath$v$}}}_{n+1} = {{\mbox{\boldmath$v$}}}^* + \dfrac{1}{\rho}\int _{t_n}^{t_{n+1}} {\rm d}s {{\mbox{\boldmath$f$}}}_{H}^\bot. 
\end{eqnarray}
The explicit form of $\int _{t_n}^{t_{n+1}} {\rm d}s {{\mbox{\boldmath$f$}}}_{H}$ is determined to approximately fulfill the rigid body condition inside the regions of the swimmer bodies and the probe particle, and it is given by 
\begin{eqnarray}
\int _{t_n}^{t_{n+1}} {\rm d}s {{\mbox{\boldmath$f$}}}_{H} &=& - \sum_{\alpha=1}^{N}  \rho \Psi_{\alpha,n+1}^{(b)} \biggl\{{{\mbox{\boldmath$v$}}}^* -\biggl[{{\mbox{\boldmath$V$}}}_\alpha^{(G)}(t_{n+1})+ {{\mbox{\boldmath$\Omega$}}}_\alpha^{(G)}(t_{n+1})\times \bigl( {{\mbox{\boldmath$r$}}}- {{\mbox{\boldmath$R$}}}_\alpha^{(G)}(t_{n+1})\bigr)  \biggr]  \biggr\} \nonumber \\
&&-   \rho \Psi_{p,n+1} \biggl\{{{\mbox{\boldmath$v$}}}^* -\biggl[{{\mbox{\boldmath$V$}}}_p(t_{n+1})+ {{\mbox{\boldmath$\Omega$}}}_p(t_{n+1})\times \bigl( {{\mbox{\boldmath$r$}}}- {{\mbox{\boldmath$R$}}}_p(t_{n+1})\bigr)  \biggr]  \biggr\}.  \label{body_force}
\end{eqnarray}
Equations (\ref{Hforce})-(\ref{HtorqueP}), and (\ref{body_force}) enforce the momentum and angular momentum exchanges between the solvent and swimmer bodies. The velocity field at the new time step is  
\begin{eqnarray}
{{\mbox{\boldmath$v$}}}_{n+1} = {{\mbox{\boldmath$v$}}}^*\bigl[1-\sum_{\alpha=1}^{N}\Psi_{\alpha,n+1}^{(b)}-\Psi_{p,n+1} \bigr] +  \sum_{\alpha=1}^N \Psi_{\alpha,n+1}^{(b)} \biggl[{{\mbox{\boldmath$V$}}}_\alpha^{(G)}(t_{n+1})+ {{\mbox{\boldmath$\Omega$}}}_\alpha^{(G)}(t_{n+1})\times \bigl( {{\mbox{\boldmath$r$}}}- {{\mbox{\boldmath$R$}}}_\alpha^{(G)}(t_{n+1})\bigr)  \biggr] \nonumber \\
+   \Psi_{p,n+1} \biggl[{{\mbox{\boldmath$V$}}}_p(t_{n+1})+ {{\mbox{\boldmath$\Omega$}}}_p(t_{n+1})\times \bigl( {{\mbox{\boldmath$r$}}}- {{\mbox{\boldmath$R$}}}_p(t_{n+1})\bigr)  \biggr]\label{velocity}
   \end{eqnarray}
with
\begin{eqnarray}
\nabla\cdot {{\mbox{\boldmath$v$}}}_{n+1}  &=&0. 
\end{eqnarray}

\end{widetext}

\section{Supporting simulations investigating front rear asymmetry in scattering of swimmers}

\begin{figure}[hbt] 
\centering
\includegraphics[width=8cm]{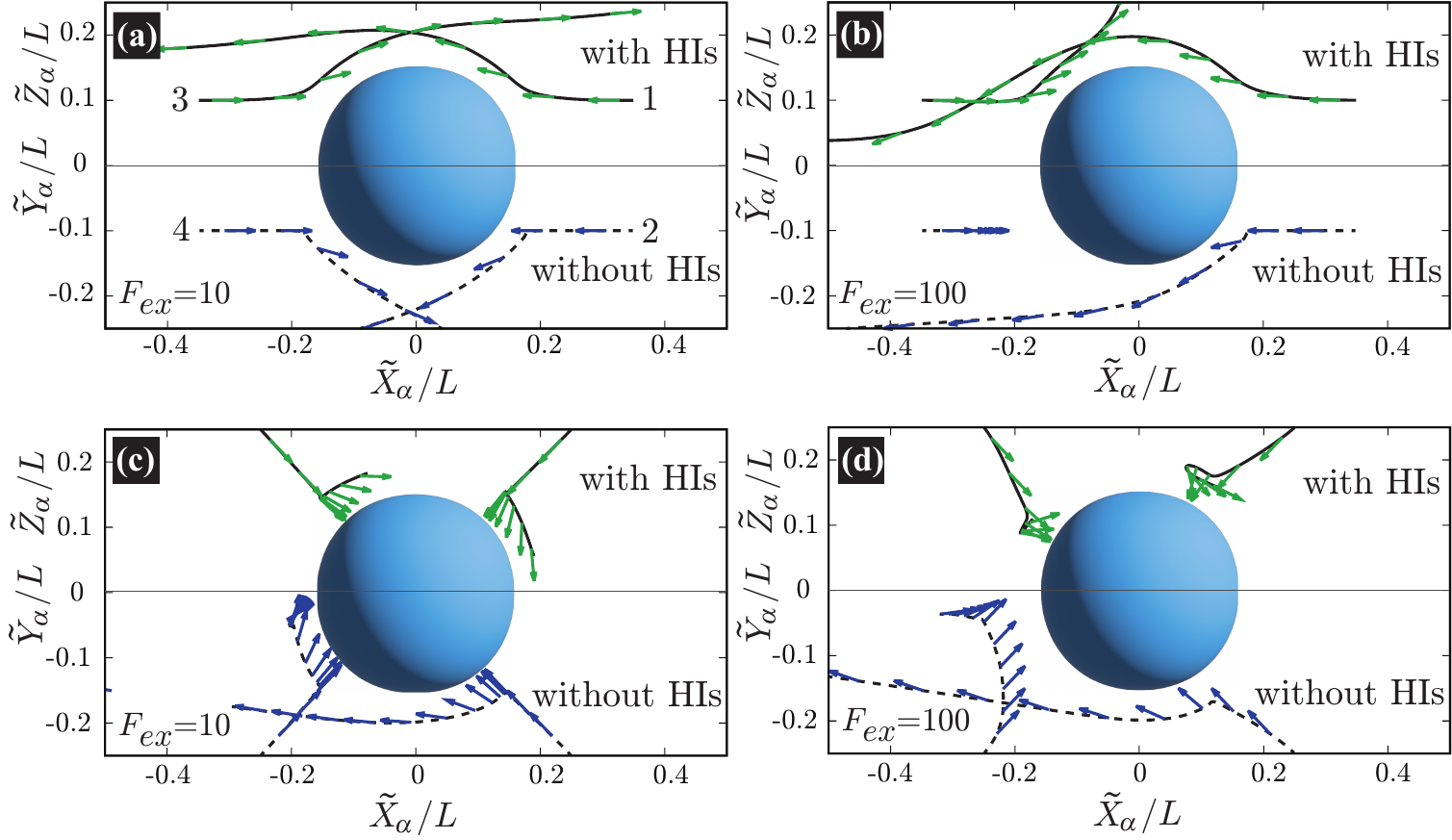}
\caption{(Color online) 
The trajectories of swimmers measured in the co-moving frame with the probe particle, ${{\mbox{\boldmath$R$}}}_\alpha^{(G)}-{{\mbox{\boldmath$R$}}}_p=({\tilde X}_\alpha,{\tilde Y}_\alpha,{\tilde Z}_\alpha)$, are plotted.  
In (a) and (b), for $F_{ex}=10$ and $100$, respectively, 
swimmers 1 and 2 are initially placed at 
${{\mbox{\boldmath$R$}}}_\alpha^{(G)} =(14a,\pm 4a,0)$ with ${\hat{\mbox{\boldmath$n$}}}_\alpha =(-1,0,0)$, 
while swimmers 3 and 4 are at 
${{\mbox{\boldmath$R$}}}_\alpha^{(G)} =(-14a,0,\pm 4a)$ with ${\hat{\mbox{\boldmath$n$}}}_\alpha =(1,0,0)$. 
On the other hand, in (c) and (d), for $F_{ex}=10$ and $100$, respectively, 
swimmers 1 and 2 are initially placed at 
${{\mbox{\boldmath$R$}}}_\alpha^{(G)} =(10a,\pm 10a,0)$ with ${\hat{\mbox{\boldmath$n$}}}_\alpha =(-\sqrt{2}/2,\mp \sqrt{2}/2,0)$,  
while swimmers 3 and 4 are at 
${{\mbox{\boldmath$R$}}}_\alpha^{(G)} =(-10a,0,\pm 10a)$ with ${\hat{\mbox{\boldmath$n$}}}_\alpha =(\sqrt{2}/2,0,\mp \sqrt{2}/2)$. 
In (a)-(d),  for swimmers 1 and 2, $({\tilde X}_\alpha/L,{\tilde Y}_\alpha/L)$ is ploted, while for swimmers 3 and 4, $({\tilde X}_\alpha/L,{\tilde Z}_\alpha/L)$ is ploted. 
By considering the spatial symmetry of the swimmer trajectories, the trajectories of swimmers 1 and 3 for the cases with HIs are plotted, and those of swimmers 2 and 4 for the cases without HIs are plotted.  
Furthermore, the arrows indicate the instantaneous directions of each swimmer.  
}
\label{FigS2}
\end{figure}

The orientation distribution of swimmers $P(\theta;x,y)$, shown in Fig. 4 in the main text, indicates that swimmers suffer from more significant scatterings at the front side than at the rear, resulting in the asymmetric shape of $P(\theta;x,y)$ between the front and rear sides. Here, we present complementary simulation results, demonstrating front rear asymmetry in scattering of swimmers with the following setup. At $t=0$, the probe particle is placed at the center of the system, and four swimmers, with active force being turned off ($F_A=0$), are arranged to face the probe in a spatially symmetric manner: swimmers 1 and 2 are on the front side and swimmers 3 and 4 are on the rear side, as illustrated in Fig. \ref{FigS2}. For $t>0$, the probe particle is dragged by a constant force ($F_{ex}$) along the $x$-direction, and the active force is turned on ($F_A=20$).  
Then, these four swimmers move toward the probe from different directions. 

The swimmer trajectories are shown in Fig. \ref{FigS2}, demonstrating that the collision conditions and the presence or absence of HIs strongly influence the trajectories. With HIs, swimmers approaching the probe particle from the rear tend to face the probe more than those approaching from the front. This behavior is attributed to the comparatively weaker repulsive nature of HIs between swimmers and the probe on the rear side than on the front. 
This repulsion on the rear side even becomes an attraction for larger values of $F_{ex}$. In contrast, without HIs, swimmers that collide with the probe from the front experience stronger repulsions than those approaching from the rear. Furthermore, as $F_{ex}$ increases, swimmers from the rear cannot keep pace with those in the front.

\section{Orientation distributions of swimmers without HIs at $F_{ex}=100$}
\begin{figure}[hbt] 
\centering
\includegraphics[width=8cm]{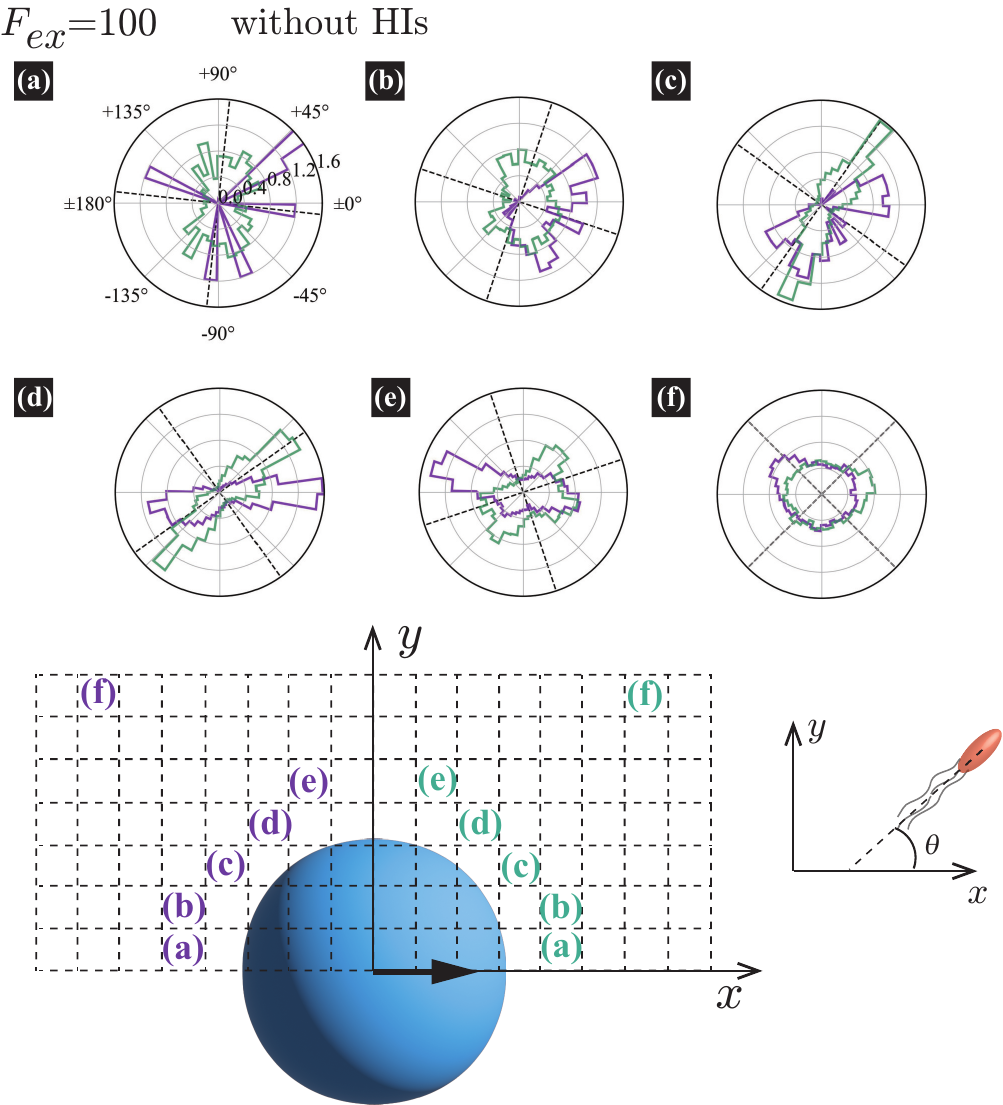}
\caption{(Color online) The orientation distributions of swimmers $P(\theta;x,y)$ 
at $R=20$, $\phi=0.032$, and $F_{ex}=100$ without HIs, with the block size being $2a=0.32R(=6.4)$.  
As described in the main text, $P(\theta;x,y)$ is defined as $P(\theta;x,y)=c(x,y)\sum_{\alpha}\langle \delta[\theta-\cos^{-1}({\hat {\mbox{\boldmath$t$}}}_\alpha\cdot {\hat {\mbox{\boldmath$x$}}})]\delta({\mbox{\boldmath$r$}}-{\mbox{\boldmath$R$}}_\alpha)\rangle$, where ${ {\mbox{\boldmath$r$}}}=(x,y,0)$, ${\hat {\mbox{\boldmath$x$}}}$ is the unit vector along the $x$-axis, and ${\hat {\mbox{\boldmath$t$}}}_\alpha={\hat {\mbox{\boldmath$n$}}}_\alpha^{||}/|{\hat {\mbox{\boldmath$n$}}}_\alpha^{||}|$ with ${\hat {\mbox{\boldmath$n$}}}_\alpha^{||}$ denoting the projected swimming direction onto the $x$-$y$ plane. 
The center of the probe sphere is located at $(x,y)=(0,0)$. 
Following Fig. 4 in the main text, the plot settings in (a)-(f) are as follows:  
The panels present the distributions at $F_{ex}=$100. 
The violet and dark-green lines correspond to $P(\theta;x,y)$ $(x<0)$ and $P(180^\circ-\theta;x,y)$ $(x>0)$, respectively, calculated in the regions indicated by identical colored characters. 
The normalization factor $c(x,y)$ is determined so that the total area enclosed by these lines equals 1. The dotted lines guide the normal and tangential directions along the probe sphere. 
}
\label{FigS3}
\end{figure}

In Fig. 4 in the main text, we present the orientation distribution of swimmers $P(\theta;x,y)$ around the probe particle at $F_{ex}=10$ and $100$ with HIs.
Here, Fig. \ref{FigS3} shows the results at $F_{ex}=100$ without HIs. 

For $x<0$, a comparison of Fig. \ref{FigS3} with Fig. 4 reveals that the swimmers tend to move along the probe surface from the rear to the front more with HIs than without HIs. This behavior is attributable to HIs between the swimmers and the probe particle; at $F_{ex}=100$ the flow field induced by the probe particle's motion is strong enough to guide the swimmers in the probe particle's moving direction. 
However, without HIs, there is a nearly equal distribution of swimmers moving from the rear to the front and in the opposite direction. On the other hand, for $x>0$, in Figs. 4 and \ref{FigS3}, we find that, in the regions labeled as (a) and (b), swimmers tend to face the probe particle more without HIs than with HIs; the swimmers are significantly scattered by hydrodynamic repulsions induced by the probe particle's movement. Also, in the regions indicated as (c)-(e), a similar distinction between with and without HIs is observed, although it is less pronounced.


\begin{thebibliography}{170}
\bibitem{comment_zeta_woHIs}
When $F_{ex}$ is sufficiently large compared to $F_A$, the impact of active forces on the resistive force acting on the probe becomes negligible, leading to a decrease in $\zeta$. 
In contrast, when $F_{ex}$ is significantly smaller than $F_A$, collision details with swimmers control the decrease in $\zeta$. In the intermediate range of $F_{ex}$, the average velocity of the probe, $\langle V_{p,x}\rangle$, becomes comparable to the swimming speed of the swimmer: for such a situation, front collisions are dominant because very few swimmers can catch up and collide from the rear side. As a result, the resistance increases with increasing $F_{ex}$. Although the observed non-monotonic $F_{ex}$ dependence of $\zeta$ without HIs itself is interesting, the present focus lies in the role of HIs. 
A detailed investigation of these behaviors will be presented elsewhere.
\end{thebibliography}
\end{document}